\documentclass[12pt]{article}   
   
\usepackage{array}   
\usepackage{epsfig}   
\usepackage{amssymb}   
\usepackage{graphics,graphpap}   
\usepackage{amssymb}   
\usepackage{amsmath}   
\usepackage[usenames]{color}  
\usepackage{slashed}   
\usepackage{graphicx}

\renewcommand{\thefootnote}{\fnsymbol{footnote}}   
   
\def\s#1{\setbox0=\hbox{$#1$}%
  \rlap{\ifdim\wd0>.7em\kern.22\wd0\else\kern.1\wd0\fi /}#1}   

   
\begin{document}   
   
\begin{titlepage}   
\begin{flushright}   
\begin{tabular}{l}   
TUM-HEP-810/11
\end{tabular}   
\end{flushright}   
\vskip1.5cm   
\begin{center}   
{\Large \bf \boldmath   A simple relation for $B_s$-mixing}   
\vskip1.3cm    
{\sc   
Alexander J. Lenz      \footnote{Alexander.Lenz@ph.tum.de}$^{,1}$}   
\vskip0.5cm   
$^1$ Physik Department, Technische Universit{\"a}t M{\"u}nchen,
D-85748 Garching, Germany
   
\vskip2cm

   
\vskip3cm   
   
{\large\bf Abstract\\[10pt]} \parbox[t]{\textwidth}{   
We reinvestigate a simple relation between 
the semileptonic CP asymmetry $a_{sl}^s$,
the decay rate difference $\Delta \Gamma_s$, 
the mass difference $\Delta M_s$ 
and $S_{\psi \phi}$
extracted from the angular analysis of the decay 
$B_s \to \psi \phi$, which is regularly used in the literature.
We find that this relation is not suited to eliminate the theory prediction for
$\Gamma_{12}$, it can, however be used to determine the size
of the penguin contributions to the decay $B_s \to \psi \phi$.
Moreover we comment on the current precision of the theory prediction for
$\Gamma_{12}$.
}   
   
\vfill   
   
\end{center}   
\end{titlepage}   
   
\setcounter{footnote}{0}   
\renewcommand{\thefootnote}{\arabic{footnote}}   
\renewcommand{\theequation}{\arabic{section}.\arabic{equation}}   
   
\newpage   
   
\section{Introduction}   
\setcounter{equation}{0}   
Currently we have some hints for deviations of experiments from standard model
predictions at the three sigma level both in the $B_d$- and the 
$B_s$-mixing system 
\cite{Lunghi:2008aa,Buras:2008nn,lnckmf,Bevan:2010gi,Lunghi:2010gv,Laiho:2011nz}.
In the $B_s$-system some of the central values differ largely 
from the corresponding standard model values - here LHCb will clearly 
tell us till the end of 2011 whether these large deviations are realized 
in nature.
In view of the expected precision of the coming data, it is mandatory
to try to achieve the same precision in the theory predictions.
In this paper we revisit a simple relation between four mixing 
observables and show that this relation can be badly violated, due to
neglecting some small quantities.
\\
After reviewing the mixing formalism in Section 2.1, we show how new physics affects 
different phases arising in the mixing of neutral B-mesons.
In Section 2.3 we derive the relation and we test it in the following section.
In Section 3 we discuss the accuracy of the theory prediction for $\Gamma_{12}$. 
Finally we conclude in Section 4.

\section{The Relation}
\subsection{Mixing formalism}
Mixing of neutral $B$ mesons is described by the off-diagonal
elements $M_{12}^q$ and $\Gamma_{12}^q$ of the mass and decay rate matrix.
The following mixing observables are determined in experiments:
the mass difference 
\begin{equation}
\Delta M_q := M_H^q - M_L^q  = 
        2 {{ |M_{12}^q|}} \left( 1 -
        { \frac{1}{8} \frac{|\Gamma_{12}^q|^2}{|M_{12}^q|^2} \sin^2 \phi_q +
...}\right),
\end{equation}
the decay width difference
\begin{equation}
\Delta \Gamma_q := \Gamma_L^q - \Gamma_H^q = 
        2 { |\Gamma_{12}^q| \cos  \phi_q }
        \left( 1 +
        { \frac{1}{8} \frac{|\Gamma_{12}^q|^2}{|M_{12}^q|^2} \sin^2 \phi_q 
        + ...}\right)
\end{equation}
and the flavour specific asymmetry (defined e.g. in \cite{NLOG12})
\begin{eqnarray}
      a_{fs}^q & = &   \mbox{Im} \frac{\Gamma_{12}^q}{M_{12}^q} { + {\cal O} \left( \frac{\Gamma_{12}^q}{M_{12}^q} \right)^2 } 
=  \frac{|\Gamma_{12}|}{|M_{12}|} \sin \left( \phi_q \right) 
    { + {\cal O} \left( \frac{\Gamma_{12}^q}{M_{12}^q} \right)^2 } \; ,       
\end{eqnarray}
where $\phi_q = \mbox{arg}( -M_{12}^q/\Gamma_{12}^q)$. A typical example of a flavor-specific decay is a
semileptonic decay, therefore this asymmetry is also called semileptonic CP-asymmetry, 
$a_{sl}^q = a_{fs}^q$. In the standard model $\Gamma_{12}^q/M_{12}^q$ is of the order of $5 \cdot 10^{-3}$, 
so one can safely neglect $(\Gamma_{12}^q/M_{12}^q)^2$.
The standard model values of these quantities were recently updated in \cite{numupdate}
using the NLO-QCD calculations from \cite{Buras:1990fn,NLOG12}.
\begin{eqnarray}
\Delta M_s^{\rm SM} =  (17.3 \pm 2.6) \; \mbox{ps}^{-1}, && \Delta M_d^{\rm SM} =  (0.54 \pm 0.09) \; \mbox{ps}^{-1},
\\
\frac{\Delta \Gamma_s^{\rm SM}}{\Gamma_s} = 0.137 \pm 0.027, && \frac{\Delta \Gamma_d^{\rm SM}}{\Gamma_d} = \left( 4.2 \pm 0.8 \right) \cdot 10^{-3},
\\
a_{fs}^{s,\rm SM} = \left(1.9  \pm 0.3 \right) \cdot 10^{-5}, && a_{fs}^{d,\rm SM} = - \left(4.1 \pm 0.6 \right) \cdot 10^{-4},
\\
\phi_s^{\rm SM} = 0.22^\circ \pm 0.06^\circ, && \phi_d^{\rm SM}   = {-4.3^\circ} \pm {1.4^\circ},
\label{phi-SM}
         \\
        A_{sl}^{b,\rm SM} = 0.506 a_{sl}^{d,\rm SM} + 0.494 a_{sl}^{s,\rm SM}       & = & (-2.0 \pm 0.3 ) \cdot 10^{-4},
\label{dimuon-SM}
         \\
        a_{sl}^{s,\rm SM} - a_{sl}^{d,\rm SM}        & = & (4.3 \pm 0.7) \cdot 10^{-4}.
\end{eqnarray}
All these values were obtained with the input parameters taken from \cite{lnckmf}.
\subsection{New Physics contributions to mixing}
In the general case of new physics being present in $B$-mixing we can write
model-indepently
\begin{eqnarray}
M_{12}^q & = & M_{12}^{q, \rm SM} \cdot \Delta_q =  
               M_{12}^{q, \rm SM} \cdot |\Delta_q | e^{i \phi_q^\Delta}\; ,
\label{M12}
\\
\Gamma_{12}^q & = & \Gamma_{12}^{q, \rm SM} \cdot \tilde \Delta_q =  
    \Gamma_{12}^{q, \rm SM} \cdot |\tilde \Delta_q | e^{- i \tilde \phi_q^\Delta}
\; \; , \; \; \tilde \Delta_q \approx 1 \; .
\label{Gamma12}
\end{eqnarray}
Then the mixing phase can be decomposed as
\begin{equation}
\phi_q = \phi_q^{\rm SM} + \phi_q^\Delta + \tilde{\phi}_q^\Delta \; .
\end{equation}
The standard model part $\phi_q^{\rm SM}$ is tiny in the case of $B_s$ mesons 
(Eq. (\ref{phi-SM})), the new 
physics contribution to $M_{12}^q$ is denoted by $\phi_q^\Delta$ and the hypothetical 
new physics contribution to $\Gamma_{12}^q$, which is strongly constrained by different 
well-measured observables is denoted by $ \tilde{\phi}_q^\Delta$. Because of these contraints
we will neglect new physics contributions to $\Gamma_{12}^q$ 
in the following.\footnote{Taking a non-negligible value of $ \tilde{\phi}_q^\Delta$
into account, will only strengthen our arguments.}
\\
A related quantity arises in the angular analysis of the decay $B_s \to \psi \phi$, which is 
sometimes confused with $\phi_s$ \cite{Lenz:2007nk}.
$ 2 \beta_s := - \mbox{arg} [(V_{tb}V_{ts}^*)^2 /(V_{cb}V_{cs}^*)^2 ]$
is the phase which appears in $b \to c \bar{c} s$ decays of neutral
B-mesons taking possible mixing into account,  so e.g. in the case $B_s \to \psi  \phi$.
$(V_{tb}V_{ts}^*)^2$ comes from the mixing (due to $M_{12}$) and 
$(V_{cb}V_{cs}^*)^2$ comes from the ratio of $b \to c \bar{c} s$ 
and $\bar{b} \to \bar{c} c \bar{s}$ amplitudes.
Sometimes $ \beta_s $ is approximated (using the PDG convention for the CKM elements!) as 
$ 2 \beta_s \approx  - \mbox{arg} [(V_{tb}V_{ts}^*)^2] \approx 
- \mbox{arg} [(V_{ts}^*)^2] $ - the error due to this approximation is on the
per mille level.
The standard model value for this angle reads \cite{lnckmf} 
\begin{equation}
2 \beta_s = \left(2.1 \pm 0.1 \right)^\circ \; .
\label{betas-SM}
\end{equation}
As mentioned above $\phi_s := \mbox{arg} [-M_{12}/\Gamma_{12}]$ is the phase that appears e.g. in
$a_{fs}^s$. In $M_{12}$ we have again the CKM elements $(V_{tb}V_{ts}^*)^2$, while we have a 
linear combination of  $(V_{cb}V_{cs}^*)^2$, $V_{cb}V_{cs}^*V_{ub}V_{us}^*$ 
and $(V_{ub}V_{us}^*)^2$ in $\Gamma_{12}$. Neglecting the latter two 
contributions - which is not justified - would yield the phase $2 \beta_s$.
Numerically $\phi_s^{\rm SM}$ is one order of magnitude smaller than $2 \beta_s$.
New physics contributions to mixing, i.e. to $M_{12}^s$ alters $\beta_s$ as
\begin{equation}
-2 \beta_s \to \phi_s^\Delta-2 \beta_s \; .
\end{equation}
Taking into account possible penguin contributions both in the standard model and beyond
one gets
\begin{equation}
-2 \beta_s + \delta_s^{\rm peng, SM} \to 
\phi_s^\Delta-2 \beta_s+ \delta_s^{\rm peng, SM}+ \delta_s^{\rm peng, NP} \; .
\label{NPinS}
\end{equation}
The penguin contributions are typically expected not to be huge, but they might easily be
of the same size as $-2 \beta_s$ \cite{pengpoll}.
An interesting exception of the decomposition in Eq.(\ref{NPinS})
is given by the standard model with four generations. In this particular model
a sizeable deviation of $-2 \beta_s$ from its standard model value is possible, see e.g.
\cite{SM4a,SM4b} for some bounds on the CKM-elements in this model.
\\
Sometimes in the literature (e.g. \cite{Grossman:2009mn,Bigi:2009df,Ligeti:2010ia})
the following quantity is used
\begin{equation}
S_{\psi \phi} = 
\sin \left( - \phi_s^\Delta + 2 \beta_s - \delta_s^{\rm peng, SM} - \delta_s^{\rm peng, NP}   \right) \; .
\end{equation}
A model independent fit \cite{lnckmf} for new physics in $B$-mixing,
gives the following currently allowed range \cite{numupdate}
\begin{equation}
S_{\psi \phi} = 0.78 ^{+0.12}_{-0.19} \; ,
\label{S_NP}
\end{equation}
which has to be compared with the SM value
\begin{equation}
S_{\psi \phi} = 0.036 \pm 0.002 \; .
\label{S_SM}
\end{equation}
Penguin contributions have been neglected in Eq.(\ref{S_NP}) and Eq.(\ref{S_SM}).
In the literature it is sometimes argued, that if the new physics contribution is sizeable, 
then we can approximate
\begin{eqnarray}
\phi_s^{\rm SM} + \phi_s^\Delta + \tilde{\phi}_s^\Delta & \approx &
\phi_s^\Delta \; ,
\label{approx1}
\\
\phi_s^\Delta-2 \beta_s+ \delta_s^{\rm peng, SM}+ \delta_s^{\rm peng, NP}
& \approx & \phi_s^\Delta \; ,
\label{approx2}
\end{eqnarray}
since the standard model phases and the possible penguin contributions are very small.
\subsection{Deriving the relation}
This approximation (Eq.(\ref{approx1}) and Eq.(\ref{approx2})) was used e.g. in 
\cite{Grossman:2009mn,Bigi:2009df,Kagan:2009gb,Ligeti:2010ia}
to derive a simple model independent relation between observables in the mixing system.
\begin{eqnarray}
a_{sl}^s & = & \left| \frac{\Gamma_{12}^s}{M_{12}^s} \right| 
               \sin \left( \phi_s^{\rm SM} + \phi_s^\Delta \right)
\nonumber \\
         & = & \frac{2 |\Gamma_{12}^s| 
                     \cos \left( \phi_s^{\rm SM} + \phi_s^\Delta \right)}
                    {2 |M_{12}^s|} 
                    \tan \left( \phi_s^{\rm SM} + \phi_s^\Delta \right)
\nonumber \\
         & = & - \frac{\Delta \Gamma_s}{\Delta M_s} 
                 \frac{S_{\psi \phi}}{\sqrt{1- S_{\psi \phi}^2}} \cdot 
                { \delta} \; ,
\label{relation-exact}
\end{eqnarray}
with the correction factor $\delta$
\begin{eqnarray}
\delta \left(\beta_s^{\rm SM},\phi_s^{\rm SM}, \delta_s^{\rm peng, SM}+ \delta_s^{\rm peng, NP}, \phi_s^\Delta  \right)
& = & \frac{\tan \left({    \phi_s^{\rm SM} }+ \phi_s^\Delta \right)}
                  {\tan \left({ -2 \beta_s^{\rm SM}}+ \phi_s^\Delta + \delta_s^{\rm peng, SM}+ \delta_s^{\rm peng, NP} \right)} \; .
\label{delta}
\end{eqnarray}
Since recent fits \cite{lnckmf,Bevan:2010gi} of the new physics contribution to the 
$B_s$ mixing phase give relatively large values
\begin{eqnarray}
\phi_s^\Delta & = &\left( -52^{+32}_{-25} \right)^\circ \; \; \mbox{at 95 \% C.L.} \; \cite{lnckmf},
\label{phasefit}
\\
\phi_s^\Delta & = &\left( -40 \pm 16      \right)^\circ \; \; \mbox{at 68 \% C.L.} \; \cite{Bevan:2010gi},
\end{eqnarray}
it seems to be justified to neglect all other contributions 
($\phi_s^{\rm SM}$, $-2 \beta_s^{\rm SM} $ and $\delta_s^{\rm peng, SM} + \delta_s^{\rm peng, NP} $) 
in Eq.(\ref{delta}) and to use $\delta = 1$. This was done often in the literature and one gets
\begin{eqnarray}
a_{sl}^s & = & - \frac{\Delta \Gamma_s}{\Delta M_s} 
                 \frac{S_{\psi \phi}}{\sqrt{1- S_{\psi \phi}^2}} \; .
\label{relation}
\end{eqnarray}
This relation corresponds 
to Eq.(23)  in \cite{Grossman:2009mn},
to Eq.(A.1) in \cite{Bigi:2009df},
to Eq.(55)  in \cite{Kagan:2009gb} and
to Eq.(3)   in \cite{Ligeti:2010ia}.
Several years earlier similar relations to Eq.(\ref{relation}) were derived 
e.g. in \cite{Blanke:2006ig,Ligeti:2006pm,Grossman:2006ce,Blanke:2006sb}.
In Eq.(7.10) of \cite{Blanke:2006ig} and Eq.(10) of \cite{Ligeti:2006pm} also
the same approximations (Eq.(\ref{approx1}) and Eq.(\ref{approx2})) 
were used, but in the final formula there was still theory input on $\Gamma_{12}/M_{12}$.
Therefore one expects also a correction factor $\tilde \delta$, which can 
be obtained from $\delta$ by replacing the {\it tangent} with the {\it sine}. Therefore $\tilde \delta$ 
deviates less from one than $\delta$. For Eq.(26) of \cite{Grossman:2006ce} only the approximation
in Eq.(\ref{approx1}) is used, which is expected to work well. For Eq.(3.53) of  \cite{Blanke:2006sb}
none of the above approximations was used.
\\
Similar relations without the approximations in Eq.(\ref{approx1}) 
and Eq.(\ref{approx2}) were presented e.g. in Eq.(10) of 
\cite{Grossman:2009mn},
Eq.(3.5) in \cite{Gedalia:2009kh} and in \cite{Kagan:2009gb}.
Kagan and Sokoloff \cite{Kagan:2009gb} also discuss deviations from 
the relation Eq.(\ref{relation}) by expanding to first order in the penguin 
contributions and the small phases, see Eq.(110-113) of \cite{Kagan:2009gb}.

\subsection{Testing the Relation}
\begin{figure}[h]
\begin{center}
\includegraphics[width=0.9\textwidth,angle=0]{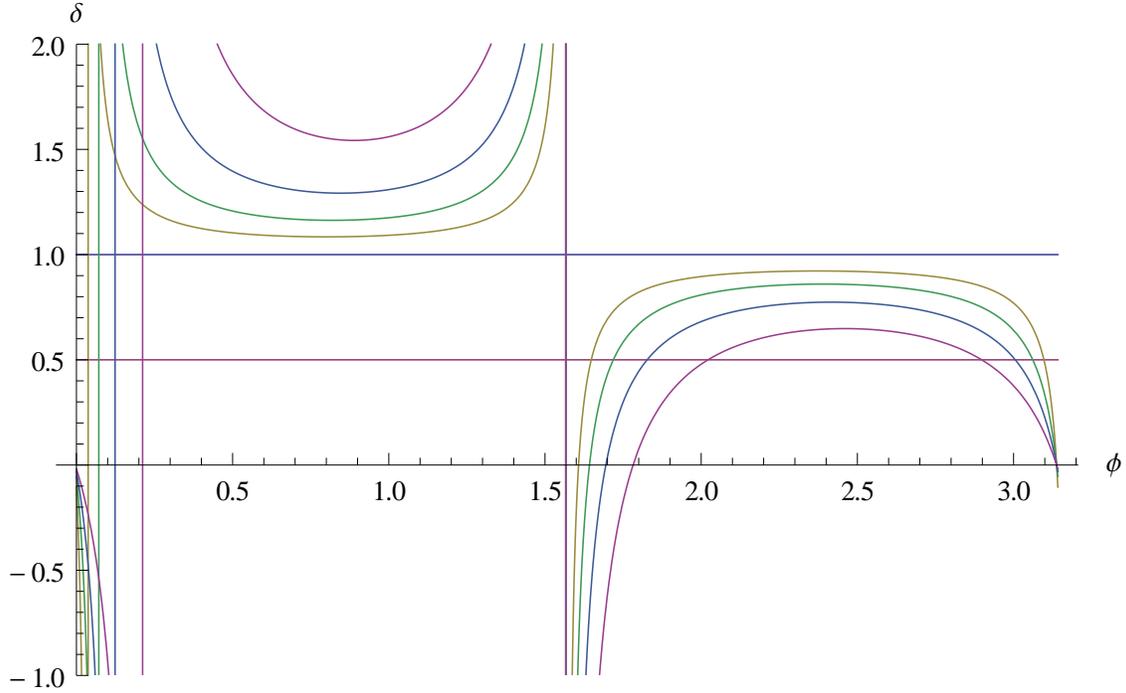}
\end{center}
\caption{Correction factor $\delta$ in dependence of the new physics phase $\phi_s^\Delta$ in $M_{12}$
for four values of the penguin contributions. The line closest to $\delta =1$ corresponds to
$\delta_s^{\rm peng, SM} + \delta_s^{\rm peng, NP} = 0^\circ$, the next to $-2^\circ$ and then $-5^\circ$ 
and $-10^\circ$.
\label{fig}}
\end{figure}

In order to test the approximations made for deriving Eq.(\ref{relation})
we plot the correction factor 
$\delta \left(\beta_s^{\rm SM},\phi_s^{\rm SM}, 
              \delta_s^{\rm peng, SM}+ \delta_s^{\rm peng, NP}, \phi_s^\Delta  
\right)$ 
in Fig. (\ref{fig}) as a function of $\phi_s^\Delta$ for four different values of
the unknown penguin contributions
$\delta_s^{\rm peng, SM} + \delta_s^{\rm peng, NP} = 0^\circ, - 2^\circ, - 5^\circ$ and $- 10^\circ$.
Comparing with  \cite{pengpoll} the second and the third values seem to be conservative and realizeable in
nature\footnote{More precise data might shrink further the allowed range for penguin contributions.}. 
$\delta_s^{\rm peng, SM} + \delta_s^{\rm peng, NP} = 10^\circ$ corresponds to large 
penguin contributions (see however \cite{Faller:2008gt}, where such a 
possibility is even not excluded within the standard model), while  
$\delta_s^{\rm peng, SM} + \delta_s^{\rm peng, NP} = 0^\circ$ corresponds to the hypothetical case that
standard model penguins and new physics penguins cancel exactly.
$\beta_s^{\rm SM}$ and $\phi_s^{\rm SM}$ are fixed to their standard model values given Eq.(\ref{betas-SM}) 
and Eq.(\ref{phi-SM}).
As expected the relation in Eq.(\ref{relation}) does not work at all 
for small new physics contributions $\phi_s^\Delta$, 
i.e. $\phi_s^\Delta \approx 0 $ or $ \phi_s^\Delta \approx \pi$, but one
can see from the plot in Fig. (\ref{fig}), that the relation in Eq.(\ref{relation}) 
does also not work for large new physics contributions.
\\
To investigate the accuracy of Eq.(\ref{relation}) further let us first discuss the case of a very large value
of the new physics mixing phase $\phi_s^\Delta$, where the approximations from 
Eq.(\ref{approx1}) and Eq.(\ref{approx2}) are expected to work best.
Taking the large $\phi_s^\Delta$-value from Eq.(\ref{phasefit}) one gets the following values for $\delta$:
\begin{displaymath}
\begin{array}{|c||c|c|c|c|}
\hline
       &\delta_s^{\rm peng, SM + NP} =   0^\circ
       &\delta_s^{\rm peng, SM + NP} = - 2^\circ
       &\delta_s^{\rm peng, SM + NP} = - 5^\circ
       &\delta_s^{\rm peng, SM + NP} = -10^\circ
\\   
\hline
\delta &  1.08638 & 1.16571 & 1.29463 & 1.54297
\\
\hline
\end{array}
\end{displaymath}
If there would be no penguin pollution at all, the relation Eq.(\ref{relation}) would
have a maximal accuracy of about $10 \%$, even for moderate penguin pollutions
Eq.(\ref{relation}) is violated by $20 \% -30 \%$ (for smaller values of $\phi_s^\Delta$, the
situation is much worse), while for large penguin contributions we have a violation
of Eq.(\ref{relation}) of more than $50 \%$. 
So even in the ideal case of a very large value of the new physics mixing phase $\phi_s^\Delta$
a violation of up to ${\cal O} (50 \%)$ can easily occur.
If $\phi_s^\Delta$ does not have the value from Eq.(\ref{phasefit}), then the relation
in Eq.(\ref{relation}) can of course be violated much stronger than  ${\cal O} (50 \%)$.
\\
Next we show a numerical example, to illustrate that a measurement of a deviation of
$\delta$ from one has to combined with theoretical input in order to obtain information 
about the phases.
Let us assume that all the quantities in Eq.(\ref{relation}) are measured and 
e.g. $\delta = 1.6$ will be found experimentally.
Such a value can correspond to very different
values of new physics phases $\phi_s^\Delta$ 
\begin{displaymath}
\begin{array}{|c||c|c|c|c|}
\hline
       &\delta_s^{\rm peng, SM + NP} =   0^\circ
       &\delta_s^{\rm peng, SM + NP} = - 2^\circ
       &\delta_s^{\rm peng, SM + NP} = - 5^\circ
       &\delta_s^{\rm peng, SM + NP} = -10^\circ
\\   
\hline
\phi_s^\Delta &   0.10457 & 0.200114& 0.352488 &0.69367
\\
\hline
\end{array}
\end{displaymath}
Without including further input (like the theoretical determination of $\Gamma_{12}$)
the new physics phase can lie somewhere between 0.1 und 0.7 and we cannot 
distinguish new physics effects in mixing from QCD effects like penguin contributions.
\\
Including however theoretical input like $\Gamma_{12}^{\rm SM}$ or $\phi_s^{\rm SM}$ we can determine
all phases, e.g.
\begin{itemize}
\item $a_{sl}^s$ from experiment and $\Gamma_{12}^{\rm SM}$ and $\phi_s^{\rm SM}$ from 
      theory allows us to extract $\phi_s^\Delta$. 
      This can then be combined with the experimental value of $S_{\psi \phi}$ or $\delta$
      to obtain $-2 \beta_s + \delta_s^{\rm Peng,SM} + \delta_s^{\rm Peng, NP}$.
      Most NP models (except e.g. the SM4) do not change the value of $-2 \beta_s$
      and therefore we can use the SM prediction for $-2 \beta_s$ to get a precise determination
      of the penguin contributions
\item The experimental values of $S_{\psi \phi}$ and $\delta$ can be used to extract
      $\phi_s^{\rm SM} + \phi_s^\Delta$. With the theoretical value of $\phi_s^{\rm SM}$
      the new physics phase $\phi_s^\Delta$ can be extracted. As above one can now extract the
      penguin contributions.
\end{itemize}
To summarize the results of this section: The relation Eq.(\ref{relation})
can be strongly violated due to the small difference between $\phi_s^{\rm SM}$ and $-2 \beta_s$
and non-vanishing penguin contributions.
Therefore a violation of Eq.(\ref{relation}), i.e. $\delta \neq 1$, does not provide 
a test of our theoretical framework for deriving $\Gamma_{12}$ and $M_{12}$.
Eq.(\ref{relation}) can not be used to eliminate the theory prediction for $\Gamma_{12}$.
The exact relation Eq.(\ref{relation-exact}) (with the correction factor $\delta$)
can however be used to extract the desired information:
Using the theory prediction for $\Gamma_{12}$ one can extract $\phi_s^\Delta$ 
and 
the size of the penguin contributions $\delta_s^{\rm peng, SM}+ \delta_s^{\rm peng, NP}$,
which is an important result.
\section{Comment on the theoretical accuracy of $\Gamma_{12}$}

The D0 collaboration measured \cite{Abazov:2010hv,Abazov:2010hj}
a very large value for the Dimuon asymmetry
\begin{equation}
 A_{sl}^{b,\rm SM} = -0.00957 \pm 0.00251 ({\rm stat}) \pm 0.00146 ({\rm syst})\, ,
\label{dimuon-EXP}
\end{equation}
which differs 3.2 $\sigma$ from the standard model prediction \cite{numupdate,NLOG12}, 
or Eq.(\ref{dimuon-SM}).
This result triggered a lot of theoretical interest, see e.g. \cite{SM4a,NP-Dimuon} 
(due to a lack of space we quoted only papers from the first two months after the
D0 result appeared).
Allowing only for new physics in $M_{12}^q$ (Eq.(\ref{M12})) one gets the following relation 
for flavor-specific/semileptonic CP-asymmetries
\begin{eqnarray}
a_{fs}^q 
&= &
{  \frac{|\Gamma_{12}^q|}{|M_{12}^{q,\rm SM}|}} 
\cdot \frac{\sin \left( { \phi_q^{\rm SM} + {
\phi^\Delta_q} }\right)}
{ |\Delta_q|}.
\end{eqnarray}
From this relation one can derive a bound on the maximal value of the dimuon-asymmetry 
\begin{equation}
 A_{sl}^{b,\rm MAX } \approx (-5 \pm 1) \cdot 10^{-3} \, ,
\end{equation}
which is about 1.5 $\sigma$ below the experimental value in Eq.(\ref{dimuon-EXP}).
\\
Due to this discrepancy (although the statistical significance is only 1.5 $\sigma$),
it was suggested \cite{Gamma_12} that new physics might also act in $\Gamma_{12}^q$, 
c.f. Eg. (\ref{Gamma12}) or that the theory prediction for $\Gamma_{12}^q$ might 
be affected by non-perturbative effects.
One possibility to circumvent hypthetical problems with $\Gamma_{12}^q$ would be the elimination
of the corresponding theory prediction with the help of Eq.(\ref{relation}),  as suggested e.g.
in \cite{Ligeti:2010ia}. But as explained above Eq.(\ref{relation}) can not be used without theory 
information on $\Gamma_{12}^q$.
\\
In order to shed some light on the necessity of the elimination of the theory prediction
for $\Gamma_{12}$ we review here its theory status.
$\Gamma_{12}^s$ has three contributions
\begin{equation}
\Gamma_{12}^s = \lambda_c^2 \Gamma_{cc} + 2 \lambda_c \lambda_u \Gamma_{uc}+ \lambda_u^2 \Gamma_{uu} \; .
\end{equation}
$\Gamma_{xy}$ corresponds to a box diagram with internal $x$- and $y-$quarks and 
$\lambda_i = V_{is}^* V_{ib}$. We investigate now the expected expansion parameter in the 
Heavy Quark Expansion for the individual $\Gamma_{xy}$.
\begin{itemize}
\item[$\Gamma_{cc}$]: This contribution dominates by far $|\Gamma_{12}|$ and $ Re (\Gamma_{12})$ and therefore
                      describes $\Delta \Gamma_s$. Since we have now two charm quarks in the intermediate
                      state, the expansion parameter of the Heavy Quark Expansion is not the inverse
                      bottom mass but a reduced mass:
                      \begin{equation}
                      \frac{\Lambda}{m_b} \to \frac{\Lambda}{\sqrt{m_b^2 - 4 m_c^2}} 
                                            = \frac{\Lambda}{m_b} \frac{1}{\sqrt{1 - 4 z}} \; .
                      \end{equation}
                      Using pole masses for the quarks one gets an expansion parameter
                      of about $ 1.3 \Lambda / m_b$. It is however well-known that the use of the
                      pole mass suffers from considerable uncertainties related to renormalons.
                      Using instead the method and parameters 
                      (MS-values at the same scale for the quark masses in $z$, which corresponds to
                       summing up logarithms of the form $z \ln z$ to all
                       orders)
                      which were used in \cite{numupdate} we get as an expansion parameter of the HQE
                      \begin{equation}
                      \frac{\Lambda}{m_b} \frac{1}{\sqrt{1 - 4 \bar z}} = (1.11 \pm 0.01) \frac{\Lambda}{m_b} \; ,
                      \end{equation}
                      that is almost identical to $\Lambda / m_b$.
                      So this simple dimensional estimate does not indicate any problems concerning
                      the convergence of the HQE. 
                      \\
                      Moreover the validity of the HQE for $\Gamma_{cc}$ can
                      be tested directly by comparing theory and experiment for $\Delta \Gamma_s$
                      and indirectly by the lifetime ratio $\tau_{B_s} / \tau_{ B_d}$, because
                      a very similar contribution arises in the theoretical determination of the
                      lifetime of the $B_s$ meson, see e.g. \cite{Lenz:2008xt}.
                      Currently no deviation from the standard model predictions are seen 
                      \cite{numupdate}, but more precise experimental numbers for $\Delta \Gamma_s$ and
                      $\tau_{B_s}$ are very desireable.
\item[$\Gamma_{uc,uu}$]: These two contributions give the dominant contribution to
                      $ Im (\Gamma_{12})$ and are therefore important for $a_{sl}$.
                      Since we now have at most one charm quark and else only light up-quarks
                      as internal quarks,
                      naive power counting shows that the HQE is given as an expansion
                      in the inverse heavy b-quark mass, which is expected to be well-behaved.
                      A very similar contribution arises in the theoretical determination of the
                      lifetime of the $B_d$ and $B^+$ mesons. Theory and experiment agree well
                      for the ratio $\tau_{B^+}/\tau_{B_d}$ \cite{numupdate}, although the theoretical
                      precision is strongly limited by a lack of knowledge of the arising 
                      non-perturbative parameters.
\end{itemize}
As we have shown, dimensional estimates do not indicate a breakdown of the convergence 
of the HQE, but instead  of dimensional estimates it is much more instructive to determine 
explicitly the size of all 
the corrections to $\Gamma_{12}$.
                      We can write
                      \begin{equation}
                      \Delta \Gamma_s = \Delta \Gamma_s^0 \left(1 + \delta^{\rm Lattice}
                                        + \delta^{\rm QCD} 
                                        + \delta^{\rm HQE} \right) \; .
                      \end{equation}
                      $\Delta \Gamma_s^0$ is the theory prediction in LO-QCD, 
                      LO-HQE (i.e. only contributions
                      of dimension 6) and with all bag parameters set to one.
                      $\delta^{\rm Lattice}$ corresponds to the deviation of the lattice 
                      results for the bag parameters from one, 
                      $\delta^{\rm QCD}$ corresponds to the NLO-QCD corrections and
                      $\delta^{\rm HQE}$ to the higher orders in the HQE.
                      With the numerical values used in  \cite{numupdate} we get the following results
                      \footnote{A detailed analysis of the error budget is given in the appendix of
                      \cite{numupdate}, we give here only the central values for the individual
                      corrections.}
                      \begin{eqnarray}
                      \Delta \Gamma_s      & = & 0.142 \mbox{ps}^{-1}
                      \\
                      \delta^{\rm Lattice} & = & - 0.14 \; ,
                      \\
                      \delta^{\rm QCD}     & = & - 0.06 \; ,
                      \\
                      \delta^{\rm HQE}     & = & - 0.19 \; .
                      \end{eqnarray}
                      All corrections are negative and smaller than $20 \% = 1/5$. So the direct
                      calculation of the first order corrections suggests that the HQE is well-behaved.
                      This can be compared with the status of 2004 \cite{Lenz:2004nx}, where considerable 
                      larger uncertainties were still present in the theory prediction for $\Gamma_{12}$.

\section{Conclusion}   
\setcounter{equation}{0}

We have investigated the relation
\begin{eqnarray}
a_{sl}^s & = & - \frac{\Delta \Gamma_s}{\Delta M_s} 
                 \frac{S_{\psi \phi}}{\sqrt{1- S_{\psi \phi}^2}} 
\end{eqnarray}
and shown that it receives verly large corrections in dependence of the value of the
new physics phase $\phi_s^\Delta$ in mixing.
Even in the {\it ideal} case of very large values of  $\phi_s^\Delta$ the relation can be violated up 
to 50 $\%$.
Therefore it can not be used to eliminate the theory prediction for $\Gamma_{12}^s$, but
including the correction factor $\delta$ from Eq.({\ref{delta})
it can be used to determine the size of the penguin contribution to the decay $B_s \to \psi \phi$,
which is a very important task. A sizeable penguin contribution can also be a signal for new
physics.
\\
We also have reinvestigated the accuracy of the theory determination of $\Gamma_{12}$ and
found no sign for unexpectedly large corrections within the framework of the HQE.
Comparision between experiment and theory predictions within the framework of the HQE shows a good 
agreement, with one exception: the central value of the dimuon-asymmetry measured by D0 is
1.5 $\sigma$ above the theory bound. 
Although this discrepancy is statistically not significant, more precise data 
for the dimuon asymmetry from TeVatron would be very helpful.
Moreover with the expected new data on $\Delta \Gamma_s$ and $\tau_{B_s}$ - in particular from LHCb -
soon much more profound conclusions about the value of $\Gamma_{12}$ can be drawn.

\section*{Acknowledgements}   
    
I thank Andrzej Buras, Alex Kagan, Uli Nierste and J{\"u}rgen Rohrwild for useful 
discussions and comments on the manuscript and Gilad Perez for challenging questions.

\end{document}